\documentclass[aps,pre,showpacs,twocolumn,floatfix,nofootinbib,showkeys]{revtex4-1}
\usepackage{amsmath,amssymb}    % need for subequations
\usepackage{graphicx}   % need for figures
\usepackage{verbatim}   % useful for program listings
\usepackage{color}      % use if color is used in text
\usepackage{subfigure}  % use for side-by-side figures
\usepackage{hyperref}   % use for hypertext links, including those to external documents and URLs
\usepackage{placeins} %stop floats moving around too much
\usepackage{amsmath}

\graphicspath{{./Figures/}}
\hypersetup{pdfborder = {0 0 0},colorlinks=true}
\newcommand{\argmax}{\operatorname*{argmax}}
\renewcommand{\d}{\text{d}} %Differential d
\newcommand{\s}{s} %Values of random variable (Small notation)
\newcommand{\sss}{\pmb{s}} %Vector of values of random variable 

\newcommand{\prop}{p_\Theta(x,\tau|y,0)} %Markov propagator
\newcommand{\p}{\hat{p}} %empirical distribution
\newcommand{\q}{q_{\Theta,\tau}} % propagated empirical distribution
\begin{document}
\title{Inferring the parameters of a Markov process from snapshots of the steady state}
\author{Simon L. Dettmer and Johannes Berg}
\email{sdettmer@thp.uni-koeln.de, and berg@thp.uni-koeln.de}
\affiliation{Institute for Theoretical Physics, University of Cologne,
  Z\"ulpicher Stra{\ss}e 77, 50937 Cologne, Germany}

\begin{abstract} 
We seek to infer the parameters of an ergodic Markov process from samples taken independently from the steady state. Our focus is on non-equilibrium processes, where the steady state is not described by the Boltzmann measure, but is generally unknown and hard to compute, which prevents the application of established equilibrium inference methods. We propose a quantity we call propagator likelihood, which takes on the role of the likelihood in equilibrium processes. This propagator likelihood is based on fictitious transitions between those configurations of the system which occur in the samples. The propagator likelihood can be derived by minimising the relative entropy between the empirical distribution and a distribution generated by propagating the empirical distribution forward in time.  Maximising the propagator likelihood leads to an efficient reconstruction of the parameters of the underlying model in different systems, both with discrete configurations and with continuous configurations. We apply the method to non-equilibrium models from statistical physics and theoretical biology, including the asymmetric simple exclusion process (ASEP), the kinetic Ising model, and replicator dynamics.
 \end{abstract}
 
 %-------------------------------------------------------------------------

%02.50.Ga Markov processes
%05.10.-a	Computational methods in statistical physics and nonlinear dynamics (see also 02.70.-c in mathematical methods in physics)
%02.30.Zz Inverse problems
%02.50.Tt	Inference methods
%89.75.-k Complex systems
%75.50.Lk Spin glasses and other random magnets
%05.70.Ln Non-equilibrium and irreversible thermodynamics


\maketitle
%%%%%%
%===========================================================
\section{Introduction}

The problem of inferring the parameters of a stochastic model from data is ubiquitous in the natural and social sciences, and engineering. Many systems, like gene regulatory networks, electric power grids, virus populations, or financial markets have a complex dynamics which is often modelled by stochastic processes. Such stochastic processes are characterised by potentially many free parameters, which need to be estimated from data. For a review in the context of the inverse Ising problem, see~\cite{Nguyen2017}. 

Here, we ask how to infer the parameters characterising a non-equilibrium stochastic process. We consider a system with configurations $x$ in some configuration space and time-homogeneous transition probabilities between configurations.
Our focus is on time-homogeneous Markov processes, which are fully defined by instantaneous transition rates. These rates are parametrized by a model with parameters denoted $\Theta$. 
Configurations can be discrete or continuous, and also time can be discrete or continuous. For the concrete example of a colloidal particle undergoing Brownian motion, the configurations $x$ are positions in space and the model parameter specifies the diffusion constant of the particle.
We restrict ourselves to ergodic processes, so for any initial state the system eventually settles into a unique steady state characterised by the steady-state probability distribution $p_\Theta(x)$.
Our aim is to infer the underlying parameters $\Theta^\text{true}$ from $M$ samples $x^\mu$, with $\mu=1,\hdots,M$ , drawn independently from the steady state distribution. 

Parameter inference hinges on the description of the empirical data by a model. For a model whose steady state $p_\Theta(x)$ is known explicitly, the maximum-likelihood estimate 
\begin{equation}
\label{maxlike}
\Theta^\text{inf} = \argmax_{\Theta} \prod_{\mu=1}^M p_\Theta(x^{\mu})
\end{equation}
provides an estimate of the model parameters which becomes exact in the limit of a large number of samples. However, for non-equilibrium models, the steady state $p_\Theta(x)$ is hard to compute and generally unknown. This is a major difference to equilibrium models and prevents the use of established inference methods.
In some cases, time series data is available and one can use the empirically observed transitions between configurations to compute the likelihood of the observed time series. This likelihood can be computed directly from the transition probabilities specified by the model; the underlying model parameters are then estimated as the parameters that maximise the likelihood of the time series~\cite{Roudi2011a,zeng2013}. Inference from time series can be performed even more efficiently using mean-field approximations~\cite{Roudi2011a,Mezard2011a,Mahmoudi2014}.

However, for many systems, classical as well as quantum, time series data is not available. An extreme case is whole-genome single-cell gene expression profiling, where cells are destroyed by the measurement process. In such cases, we have only independent samples from which to infer the model parameters. To this end, we use the transition rates between configurations and their dependence on the model parameters to construct a quantity we call the propagator likelihood. We show how this likelihood can be used to infer the model parameters from independent samples taken from the steady state.

This article is organised as follows: First, we introduce the propagator likelihood through an intuitive argument and then offer a systematic derivation based on relative entropies. Second, we apply the propagator likelihood to pedagogical examples with both discrete and continuous configurations, specifically the asymmetric simple exclusion process (ASEP) and the Ornstein-Uhlenbeck process. Finally, we address the more challenging problem of inferring the parameters of two prominent models from statistical physics and theoretical biology: the kinetic Ising model and replicator dynamics.

%===========================================================
\section{The propagator likelihood}
Suppose we knew the functional dependence of the steady-state distribution $p_\Theta(x)$ on the model parameters $\Theta$. Then a standard approach would be to maximise the (log-)~likelihood of the samples
\begin{align}
\mathcal{L}(\Theta)  = \frac{1}{M} \sum_{\mu=1}^M \log p_\Theta(x^\mu)  = \sum_x \hat{p}(x) \log p_\Theta(x) \ ,
\label{eq:log-likelihood}
\end{align}
where the set of sampled configurations characterises the empirical distribution $\hat{p}(x)$ with probability mass function
\begin{equation}
\hat{p}(x)=\frac{1}{M}\sum_{\mu=1}^M \delta_{x^\mu,x} \ ,
\label{eq:sampled-dist-def}
\end{equation}
and $\delta_{x^\mu,x}$ denotes a Kronecker-$\delta$. 

However, in non-equilibrium systems we frequently do not know the steady-state distribution. Non-equilibrium systems lack detailed balance, so the steady state is not described by the  Boltzmann distribution and lacks a simple characterisation. Our solution to this inference problem is based on exploiting one elementary fact: since the distribution $p_\Theta$ is stationary, it remains unchanged if we propagate it forward in time by an arbitrary time interval. Thus, we can replace the steady-state distribution $p_\Theta(x)$ in the log-likelihood function \eqref{eq:log-likelihood} with the same distribution propagated forward in time $\sum_y  \prop p_\Theta(y)$. 
The propagator $\prop$ is the conditional probability of observing the system in configuration $x$ at time $t=\tau$, given it was in configuration $y$ at time $t=0$. 
By replacing the unknown steady-state distribution $p_\Theta(y)$ with the empirical distribution $\hat{p}(y)$, we arrive at the propagator likelihood
\begin{align}
\mathcal{PL}(\Theta;\tau)&=\sum_x \hat{p}(x) \log \sum_y \prop \hat{p}(y) \notag \\
&=\frac{1}{M} \sum_{\mu=1}^M \log \left(\frac{1}{M}\sum_{\nu=1}^M p_\Theta(x^\mu,\tau |x^\nu,0)\right) \ .
\label{eq:propagator-likelihood-def}
\end{align}
In this way, we have shifted the parameter-dependence from the (unknown) steady-state distribution $p_\Theta(x)$ to the (known) propagator $p_\Theta(x,\tau |y,0)$. For models with continuous configurations, $p_\Theta(x^\mu,\tau |x^\nu,0)$ is the transition probability density. 
The propagator likelihood~\eqref{eq:propagator-likelihood-def} has a straightforward probabilistic interpretation:
$\frac{1}{M}\sum_{\nu=1}^M p_\Theta(x,\tau |x^\nu,0)$ is a probability distribution over $x$, conditional on the sampled configurations $\{x^\nu\}$. The propagator likelihood is the corresponding log-likelihood of this probability distribution, evaluated for $M$ independent draws of the empirically observed configurations and rescaled by $M$. 
In the limit $\tau \to \infty$, the propagator likelihood~\eqref{eq:propagator-likelihood-def} approaches the log-likelihood~\eqref{eq:log-likelihood}, since $\lim_{\tau\rightarrow \infty}p_\Theta(x,\tau|y,0)\equiv p_\Theta(x)$. However, the complexity of calculating the propagator increases with $\tau$.

In principle, the propagation time interval $\tau$ is arbitrary: it parameterizes different measures of how close a given empirical probability distribution is to being stationary under a particular set of model parameters. Different choices of $\tau$ will be discussed in sections~\ref{sec:OUP} and~\ref{sec:kineticIsing}. 
Maximizing the propagator likelihood does not involve sampling the probability distribution at different times, but seeks model parameters that would leave the empirical distribution invariant, if one did propagate it forward in time.
Correspondingly, although rates of transitions between configurations $x^\nu$ and $x^\mu$ appear in~\eqref{eq:propagator-likelihood-def}, these transitions are entirely fictitious: the empirical configurations $\{x^\nu\}$ are sampled independently from the non-equilibrium steady state. 
In the following, we will assume that the parameters maximizing the propagator likelihood  are unique. One might want to prove this for the particular model used by analytically calculating the propagator likelihood and checking its convexity.

\begin{figure}[h]
\center
\includegraphics[width=0.5\textwidth]{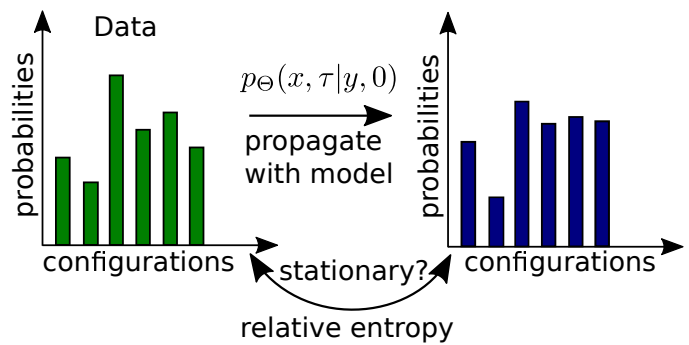}
\caption{The propagator likelihood. The set of independent samples $\{x^\mu\}_{\mu=1}^M$ characterize the empirical distribution $\hat{p}$ defined by equation~\eqref{eq:sampled-dist-def}, shown on the left. We use the transition probabilities $\prop$ to propagate $\hat{p}$ forward in time by an arbitrary interval $\tau$ to generate a new distribution $\q$ (see Eq.\eqref{eq:d-def}), shown on the right. The functional form of the propagator is thought to be known, but it is parametrized by a set of unknown parameters $\Theta$. Demanding stationarity of the empirical distribution, we can estimate the underlying parameters $\Theta^\text{true}$ by finding the parameters  $\Theta^\text{inf}$ that minimise the distance between $\p$ and $\q$ as measured with relative entropy. This is equivalent to maximising the propagator likelihood (see main text).}
 \label{fig:concept}
\end{figure}
%-----------------------------------------------------------------------------------------------------
\subsection{Minimising relative entropy}

A second interpretation of the propagator likelihood can be found by rephrasing parameter inference from a steady state as finding a set of parameters $\Theta$ such that the propagator $p_\Theta(x,\tau |y,0)$ is compatible with the empirical distribution $\hat{p}$ being stationary~(see Fig.~\ref{fig:concept}). Demanding stationarity corresponds to requiring that $\hat{p}$ is in some sense close to a distribution $q_{\Theta,\tau}$ generated by propagating the empirical distribution for an arbitrary time interval $\tau$,
\begin{equation}
 q_{\Theta,\tau}(x)=\sum_y p_\Theta(x,\tau |y,0) \hat{p}(y) \ .
 \label{eq:d-def}
\end{equation}
To quantify this notion of closeness for discrete configurations, we use the relative entropy or Kullback-Leibler divergence~\cite{Kullback1951}
\begin{equation}
D(\hat{p} \| q_{\Theta,\tau})=\sum_x \hat{p}(x) \log \frac{\hat{p}(x)}{q_{\Theta,\tau}(x)} \ .
\label{eq:KL}
\end{equation}
Inserting the probability mass function $q_{\Theta,\tau}(x)$ defined by~\eqref{eq:d-def} into the relative entropy, we find that the relative entropy can be written as the negative sum of the Shannon entropy of the empirical distribution, $S(\hat{p})=-\sum_x \hat{p}(x) \log \hat{p}(x)$ and the propagator likelihood~\eqref{eq:propagator-likelihood-def}:
\begin{equation}
D(\hat{p} \| q_{\Theta,\tau})=-S(\hat{p})-\mathcal{PL}(\Theta;\tau) \ .
\label{eq:KL-decomposition}
\end{equation}
The first term depends only on the sampled configurations and is independent of the model parameters; thus minimising the relative entropy with respect to $\Theta$ is equivalent to maximising the propagator likelihood. Furthermore, due to the positivity of relative entropy, the propagator likelihood is bounded from above by the negative Shannon entropy, and this bound will be saturated  only for a model that makes the empirical distribution exactly stationary. The propagator likelihood~\eqref{eq:propagator-likelihood-def} thus emerges from a variational approach aiming to find the model parameters that are most consistent with the sampled distribution being the steady state.

A similar argument can be made also for models with continuous configurations $x\in\mathbb{R}^d$. We consider the differential relative entropy $D=\int \d x \ \hat{p}_s(x) \log (\hat{p}_s(x)/q_{\Theta,\tau,s}(x))$, which can be computed by estimating the probability density of the steady state from the samples via a Gaussian mixture model $\hat{p}_s(x)=\frac{1}{M}\sum_{\mu=1}^M \exp(-(x-x^{\mu})^2/2s^2)/(2\pi s^2)^{d/2}$. Here, $s>0$ is the width of the Gaussians in the mixture model, and $q_{\Theta,\tau,s}(x)=\int \d y \ p_{\Theta}(x,\tau|y,0) \hat{p}_s(y)$ denotes the time-propagated density estimate. Minimising this estimate of the differential relative entropy is then equivalent to maximising a quantity that converges to the propagator likelihood for $s\rightarrow 0$.

It is easy to show that the maximum propagator likelihood estimate $\Theta^\text{inf}$ converges to the underlying parameters  $\Theta^\text{true}$ in the limit of large sample sizes: for $M\rightarrow \infty$, the empirical distribution $\hat{p}(y)$ converges to the steady-state distribution $p_{\Theta^\text{true}}(y)$. Hence, the propagator likelihood converges to $\mathcal{PL}(\Theta;\tau,{M=\infty})=\sum_xp_{\Theta^\text{true}}(x) \ln \sum_y p_\Theta(x,\tau |y,0) p_{\Theta^\text{true}}(y)$. According to~\eqref{eq:KL-decomposition}, this function has its maximum over $\Theta$ where the relative entropy between the underlying distribution $p_{\Theta^\text{true}}(x)$ and its propagated version $\sum_y p_\Theta(x,\tau |y,0) p_{\Theta^\text{true}}(y)$ is minimal. This minimum is realised for $\Theta=\Theta^\text{true}$, since the relative entropy is non-negative and the steady-state by definition remains unchanged when propagated with the parameter value $\Theta=\Theta^\text{true}$.

%==========================================================
\FloatBarrier
\section{Models with discrete configurations}

%---------------------------------------------------------------------------------------------------------------------
\subsection{Discrete time: a simple two-configuration model}
\begin{figure}[hbtp!]
\includegraphics[width=0.45\textwidth]{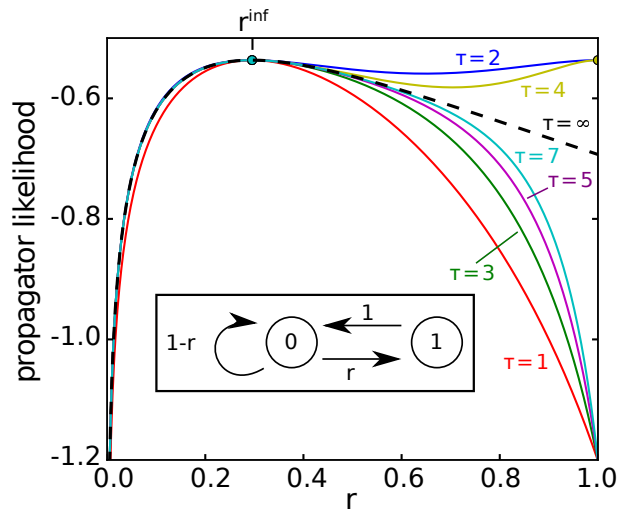}
\caption{The propagator likelihood for a simple two-configuration system. The inset shows the single-step dynamics of the system with configurations 0 and 1, controlled by the hopping probability $r\in(0,1)$. In the main figure, the solid lines show the propagator likelihood for varying propagation time intervals $\tau$. The dashed line shows the log-likelihood \eqref{eq:log-likelihood}, which corresponds to an infinite propagation time interval. The maximum likelihood estimate of the hopping probability, ${r}^\text{inf}=\frac{1-\hat{p}(0)}{\hat{p}(0)}$, is marked on the top axis and coincides with the maximum for all propagator likelihoods with an uneven number of time steps $\tau$ (see the main text for the case of even time steps).}
\label{fig:two-state}
\end{figure}

To illustrate the propagator likelihood with a toy example, we consider a system with only two configurations, denoted by $0$ and $1$ (see inset of Fig.~\ref{fig:two-state}). At each time step, if the system is in configuration $1$, it moves to configuration $0$. If it is in configuration $0$, it moves to configuration $1$ with probability $r\in(0,1)$ or remains in configuration $0$ with probability $1-r$. The steady-state distribution is easily computed, giving $p_r(0)=1/(1+r)$ and $p_r(1)=1-p_r(0)=r/(1+r)$. 

We are now given samples $\{x^\mu\}_{\mu=1}^M\in \{0,1\}^M$ taken independently from the steady state  and want to infer the model parameter $r$. The empirical distribution is given by the frequencies of the two configurations, $\hat{p}(0)=\frac{1}{M}\sum_{\mu=1}^M \delta_{0,x^\mu}$ and $\hat{p}(1)=1-\hat{p}(0)$. Since we know the steady state for this particular model, we can infer $r$ from the relationship $\langle\hat{p}(0)\rangle=1/(1+r)$, yielding $r^\text{inf}=(1-\hat{p}(0))/\hat{p}(0)$. For comparison, we also use the propagator likelihood~\eqref{eq:propagator-likelihood-def} with the single-step propagator $p_r(x,\tau=1|y,0)=\delta_{y,1}\delta_{x,0}+\delta_{y,0}(r\delta_{x,1}+(1-r)\delta_{x,0})$, giving
\begin{align}
\mathcal{PL}(r;1)&=\hat{p}_0 \log (\underbrace{(1-r)\hat{p}_0}_{0\rightarrow 0} +\underbrace{\hat{p}_1}_{1\rightarrow 0})+\hat{p}_1\log(\underbrace{r\hat{p}_0}_{0\rightarrow 1}) \notag \\
&=\hat{p}_0\log(1-r\hat{p}_0)+(1-\hat{p}_0)\log(r\hat{p}_0) \ .
\end{align}

Maximising the propagator likelihood analytically with respect to $r$ by setting $\frac{\d \mathcal{PL}}{\d r}(r^\text{inf})=0$, we recover the same result as obtained above by analysing the known steady-state distribution. Indeed, for uneven propagation time intervals, the propagator likelihood shows a unique maximum at the same point where the likelihood has its maximum, $r^\text{inf}=\frac{1-\hat{p}_0}{\hat{p}_0}$. Also, the propagator likelihood approaches the log-likelihood for increasing $\tau$, as expected. For even propagation time intervals, however, a second (global) maximum occurs at the boundary $r=1$: since the choice $r=1$ makes the two configurations simply exchange their probabilities in each step, the Markov chain loses its ergodicity and becomes periodic.  In this case, \emph{any} distribution is stationary over an even number of time steps.
While stationarity with respect to a single time step is both necessary and sufficient to define the steady state, stationarity with respect to longer propagation time intervals is necessary but not sufficient. Hence, spurious maxima of the likelihood can appear when using longer propagation time interval. 
%For this reason, in systems with discrete time, we will use the propagator likelihood with propagation time interval $\tau=1$.

%---------------------------------------------------------------------------------------------------------------------
\subsection{Continuous time: the asymmetric simple exclusion process (ASEP)}
Markov processes with discrete configurations in continuous time are characterised by instantaneous transition rates between distinct configurations $W_\Theta(x|y)=\lim_{\tau \rightarrow 0} p_\Theta(x,\tau|y,0)/\tau \ , \ (x\neq y)$. The system hops away from configuration $y$ at a random time that is exponentially distributed with parameter $-W_\Theta(y|y)\equiv\sum_{x\neq y}W_\Theta(x|y)$. For the purpose of inferring the model parameters, it is convenient to map the continuous-time process onto a discrete-time process with the same steady state. This can be achieved by choosing the single-step transition matrix
\begin{equation}
\tilde{p}_\Theta(x,\tau=1|y,0)=\delta_{x,y}+\lambda W_\Theta(x|y) \ .
\label{eq:discrete_time_version}
\end{equation}
The parameter $\lambda$ affects the overall rate at which transitions occur. Choosing $0<\lambda<[\max_{y} \{-W(y|y)\}]^{-1}$ ensures a well-defined stochastic matrix. Since the steady-state distribution $p_{\Theta}(y)$ itself is not associated with a time scale, the choice of $\lambda$ is in principle arbitrary.

\begin{figure}[h!]
\includegraphics[width=0.45\textwidth]{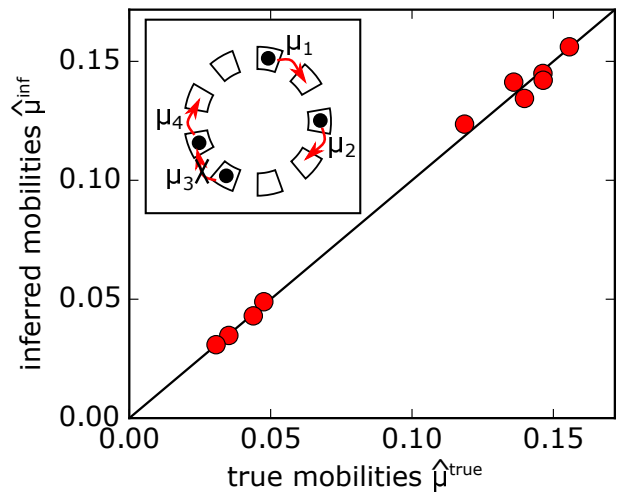}
\caption{Reconstruction of hopping rates in the asymmetric simple exclusion process (ASEP). The inset schematically shows the dynamics: $K$ particles move on a periodic one-dimensional lattice with $N>K$ lattice sites, see text. In the main figure, we plot the  relative mobilities $\hat{\mu}^\text{inf}_i$ inferred using the propagator likelihood versus the underlying relative mobilities $\hat{\mu}^\text{true}_i={\mu}^\text{true}_i/\sum_j {\mu}^\text{true}_j$ that were used to generate the data. We simulated $K=10$ particles hopping on a lattice with $N=15$ sites and took $M=10^{10}$ samples independently from the steady state. The underlying mobilities $\mu_i$ were drawn independently from a uniform distribution on the unit interval $(0,1)$.}
\label{fig:exclusion}
\end{figure}

As an example of a model with continuous time, we consider the asymmetric simple exclusion process (ASEP) on a ring with asynchronous updates (see inset of Fig.~\ref{fig:exclusion}). The ASEP is a simple model of a driven lattice gas and has been applied to traffic flow, surface growth, and directed paths in random media~\cite{Krug1996,Evans1997,Derrida1998}. 

The steady-state distribution in 1D can be calculated analytically in terms of matrix products~\cite{Derrida1998,Evans1997}. In higher dimensions, however, there is no such systematic approach and, to the best of our knowledge, the steady-state distribution is unknown. 

The model consists of $K$ particles moving on a periodic one-dimensional lattice with $N>K$ lattice sites. Each lattice site can be occupied by at most one particle. Particles labelled $i=1,\ldots,K$ independently attempt to jump one step in the clockwise direction at a rate $\mu_i$, which is called the intrinsic mobility or hopping rate of a particle.
The configuration of the system can be characterised by the number of free lattice sites in front of each particle, $\mathbf{n}=(n_1,\hdots,n_K)\subset (\mathbb{N}_0)^K$, with the restriction that the particle gaps add up to the number of free lattice sites: $n_1+n_2+\hdots+n_K=N-K$. 
For the transition $\mathbf{n}=(n_1,\hdots,n_K)\rightarrow \mathbf{n}'=(n'_1,\hdots,n'_K)$ between two distinct configurations there is a non-zero transition rate only if the configurations are connected by the jump of a single particle $i$, \textit{i.e.} all gaps are identical except for the gap in front of particle $i$, which must be decreased by one, $n'_i=n_{i}-1$, and the gap behind particle $i$, which must by increased by one $n'_{i-1}=n_{i-1}+1$. The transition rate is then simply the hopping rate of the particle $W_{\boldsymbol{\mu}}(\mathbf{n}'|\mathbf{n})={\mu}_i$. 
To infer the parameters, we define a discrete-time version of the process with transition probabilities defined by \eqref{eq:discrete_time_version}. We choose $\lambda$ such that the hopping rates add to one, $\lambda=(\mu_1+\mu_2+\hdots+\mu_K)^{-1}$ in \eqref{eq:discrete_time_version}. The steady-state distribution is characterised by the relative hopping rates $\hat{\mu}_i \equiv \mu_i/\sum_j \mu_j$. The single-step propagator likelihood of the discrete-time process then reads
\begin{align}
\mathcal{PL}(\hat{\boldsymbol{\mu}},1)=\sum_{\mathbf{n}'} \hat{p}(\mathbf{n}') \log \left\{ \hat{p}(\mathbf{n}')+\sum_{\mathbf{n}}  W_{\hat{\boldsymbol{\mu}}}(\mathbf{n}'|\mathbf{n})\hat{p}(\mathbf{n})\right\} \ .
\end{align}

We use this result to evaluate the propagator likelihood~\eqref{eq:propagator-likelihood-def} and infer the relative mobilities $\hat{\mu}_i$. As an example, we consider a system of $K=10$ particles hopping on $N=15$ lattice sites. The particle mobilities $\mu_i$ are independently and uniformly drawn from the interval $(0,1)$. We generate $M=10^{10}$ Monte Carlo samples, recorded every $10$ jumps after an initial settling time interval of $10^5$ jumps to reach the steady state. We then maximise  the propagator likelihood numerically using the sequential least squares programming algorithm as implemented in the SciPy library~\cite{SciPy}. In Fig.~\ref{fig:exclusion} we plot the inferred relative mobilities versus the relative mobilities used to generate the samples.

%===========================================================
\FloatBarrier
\section{Models with continuous configurations}

Markov processes with continuous configurations pose an additional challenge: Finite-time propagators are generally not known explicitly. Instead, finite-time propagators are characterised indirectly as the solution of a Fokker-Planck equation. Rather than solving a Fokker-Planck equation, which for systems with a large number of degrees of freedom is generally infeasible, we proceed by approximating the propagator for short time intervals $\tau$ via a linearisation of the corresponding Langevin equation (LE) that describes the stochastic dynamics of the model.

Again, we first demonstrate this procedure using a toy model. We consider one of the simplest processes with continuous configurations, the Ornstein-Uhlenbeck process (OUP), which describes the Brownian dynamics of an overdamped particle in a quadratic potential. Note that, again, for this particular case the steady-state distribution is known exactly, so one could infer the model parameters using the standard maximum likelihood approach. We use this case to illustrate the propagator likelihood before turning to more complex models where the likelihood-based approach is not feasible. 
%---------------------------------------------------------------------------------------------------------------------
\subsection{The Ornstein-Uhlenbeck process}
\label{sec:OUP}
Consider a single particle diffusing in a one-dimensional harmonic potential $U(x)=\frac{b}{2}x^2$ with diffusion constant $\sigma^2$. 
A physical realisation of this model is a colloidal particle in solution being held in place by optical tweezers and confined to a one-dimensional channel. The dynamics of the particle is modelled by the Langevin equation
\begin{equation}
\frac{\d x}{\d t}= -b x +\sigma \xi(t)  \ ,
\label{eq:ou-langevin}
\end{equation}
where the random force $\xi(t)$ describes $\delta$-correlated white noise interpreted in the It{\^o} convention.

\begin{figure}[hbtp]
\includegraphics[width=0.45\textwidth]{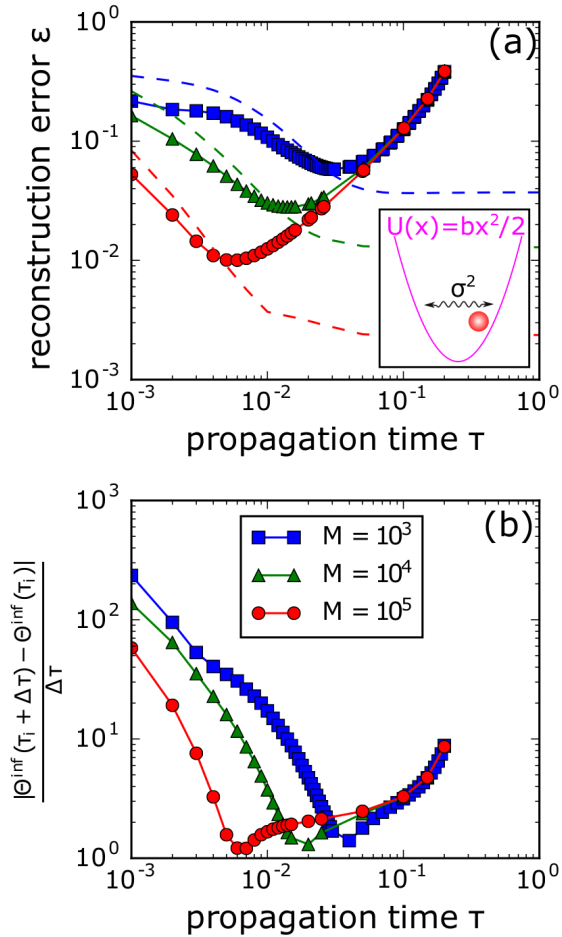}
\caption{Parameter inference in the Ornstein-Uhlenbeck process. (a) The inset shows a schematic plot of the model describing a single particle moving in the harmonic potential $U(x)=bx^2/2$. In the main figure, we show the relative reconstruction error $\epsilon=|\Theta^\text{inf}-\Theta^\text{true}|/\Theta^\text{true}$ of the parameter $\Theta=b/\sigma^2$ (characterising the steady state) versus the dimensionless propagation time interval $\tau$ used in the propagator for sample sizes $M=10^3$ ($\blacksquare$), $M=10^4$ ($\blacktriangle$), and $M=10^5$ ($\bullet$). The solid lines with markers show the reconstruction errors for the approximate short-time propagator, the dashed lines indicate the reconstruction errors for the exact finite-time propagator. \newline
(b) shows the estimated rate of change of the inferred parameter with respect to the propagation time interval $\tau$. The rates of change are computed using forward difference quotients $|\partial \Theta^\text{inf}/\partial \tau(\tau_i)|\approx |\Theta^\text{inf}(\tau_{i}+\Delta \tau)- \Theta^\text{inf}(\tau_{i})|/\Delta \tau$ and are shown on the vertical axis for the differentiation step size $\Delta \tau=10^{-3}$. The minimal rate of change corresponds to the optimal choice of the propagation time interval (see main text).\newline
The data was generated by independent sampling from the stationary distribution, \textit{i.e.} a centred Gaussian with variance $\sigma^2/(2b)=1/4$. In order to remove fluctuations between different sample sets $\{x_\mu\}_{\mu=1}^M$ and demonstrate the dependence of the average error on the sample size $M$ and propagation time interval $\tau$, the results were averaged over 50 independent sample sets. The minima of the reconstruction error and the rate of change coincide also for individual sample sets, while the position of the minima may vary across sample sets. }
\label{fig:ornstein-uhlenbeck}
\end{figure}

As for the exclusion process, one model parameter must be eliminated by rescaling time, since the steady-state distribution is time-independent. We rescale time to be dimensionless with $t'=t \sigma^2$, so that the particle has unit diffusivity. To calculate the propagator likelihood for short time intervals $\tau \ll 1$, we linearise the LE~\eqref{eq:ou-langevin} in time
\begin{equation}
x(\tau)\approx x(0)-\frac{b}{\sigma^2} x(0) \tau +\int_0^{\tau} \d t' \xi(t') \ .
\label{eq:ou-langevin-linearized}
\end{equation}
Since the integrated white noise $\int_0^{\tau} \d t' \xi(t')$ is normally distributed with mean 0 and variance $\tau$, we obtain an approximate short-time Gaussian propagator
\begin{equation}
p_{b/\sigma^2}(x,\tau|y,0) \approx \frac{\exp\left(-[x-\overline{x}]^2/2\tau\right)}{\sqrt{2\pi \tau}}  \  ,
\label{eq:ou-approximate-propagator}
\end{equation}
where $\overline{x}=y-(b/\sigma^2)y\tau$ is the most likely future position of the particle.

Such a Gaussian form of the propagator emerges for any linearised LE with white noise and is not specific to the OUP. For coloured and multiplicative noise, $\xi(t)\rightarrow f(x(t),t)\eta(t)$, where $f$ is some function and the random force $\eta(t)$ has a finite correlation time, we can proceed similarly. In this case, the normal distribution of the integrated white noise is replaced with the appropriate distribution of the integrated coloured noise ${\int_0^{\tau} \d t' f(x(t'),t')\eta(t') \approx f(x(0),0)\int_0^{\tau} \d t' \eta(t')}$.

Inserting the short-time propagator~\eqref{eq:ou-approximate-propagator} into the propagator likelihood~\eqref{eq:propagator-likelihood-def}, we perform a one-dimensional maximisation of the propagator likelihood to infer the parameter $\Theta=b/\sigma^2$. Fig.~\ref{fig:ornstein-uhlenbeck}(a) shows the relative reconstruction error versus the dimensionless propagation time interval $\tau$ for various sample sizes, both for the short-time propagator and for the exact finite-time propagator. The non-monotonic behaviour of the error for the short-time propagator shows that the optimal choice for $\tau$ involves a trade-off: At short time intervals $\tau$, the distances typically crossed during the interval $\tau$ are small. In this case, the sum over pairs of sampled configurations in the propagator likelihood~\eqref{eq:propagator-likelihood-def} is dominated by few transitions with small steps, and, in the limit $\tau \to 0$ it is dominated by transitions of the type $x^{\mu} \to x^{\mu}$.  
For this reason, the parameter inference at small values of $\tau$ is more strongly affected by sampling fluctuations than at large values of $\tau$. At large values of $\tau$, on the other hand, the approximation used to derive the short-time propagator \eqref{eq:ou-approximate-propagator} becomes invalid. As a result, both the optimal value of $\tau$ and the total reconstruction error decrease as the sample size is increased.

The exact finite-time propagator exhibits only sampling fluctuations, so the reconstruction error decreases monotonically with $\tau$, converging to the maximum likelihood estimate at large $\tau$. Note that the results for the approximate and exact propagators do not converge for $\tau \rightarrow 0$, since the \emph{relative difference} of the propagators converges to $0$ only for the peak $x=y$, even though the \emph{absolute difference} converges to $0$ for all values of $x$.

\paragraph{Choosing the optimal propagation time interval.}
The non-monotonic behaviour of the reconstruction error $\epsilon=|\Theta^\text{inf}-\Theta^\text{true}|/\Theta^\text{true}$ raises the question how to choose the optimal propagation time interval without prior knowledge of the underlying parameter $\Theta^\text{true}$. We find an answer by assuming that the error is a smooth function of the propagation time interval: we seek the minimal error by demanding $0=\partial \epsilon/\partial \tau =\frac{\text{sgn}( \Theta^\text{inf}- \Theta^\text{true})}{|\Theta^\text{true}|} \frac{\partial \Theta^\text{inf}}{\partial \tau} \sim \partial \Theta^\text{inf}/\partial \tau$. The error derivative will become small only for $\partial \Theta^\text{inf}/\partial \tau \rightarrow 0$.
The latter quantity can be estimated directly from the data by repeating the inference for a set of propagation time intervals $\{(\tau_i,\tau_i+\Delta \tau)\}$ and computing the forward difference quotients $\partial\Theta^\text{inf}/\partial \tau(\tau_i)\approx[\Theta^\text{inf}(\tau_{i}+\Delta \tau)- \Theta^\text{inf}(\tau_{i})]/\Delta \tau$. Since estimating the derivative from the data will incur numerical errors, we relax the condition $0=\partial\Theta^\text{inf}/\partial \tau$ and demand only that $|\partial\Theta^\text{inf}/\partial \tau|$ is minimal. In Fig.~\ref{fig:ornstein-uhlenbeck}(b) we show that these minima indeed coincide with the optimal choice of $\tau$ as judged from the reconstruction error shown in Fig.~\ref{fig:ornstein-uhlenbeck}(a).

%===========================================================
\FloatBarrier
\section{Non-equilibrium models in statistical physics and theoretical Biology}

We now turn to non-equilibrium applications where the standard maximum likelihood approach is not feasible, as the steady-state distribution is unknown. 

%------------------------------------------------------------------------------
\subsection{The kinetic Ising model}
\label{sec:kineticIsing}

\begin{figure}[h]
\includegraphics[width=0.45\textwidth]{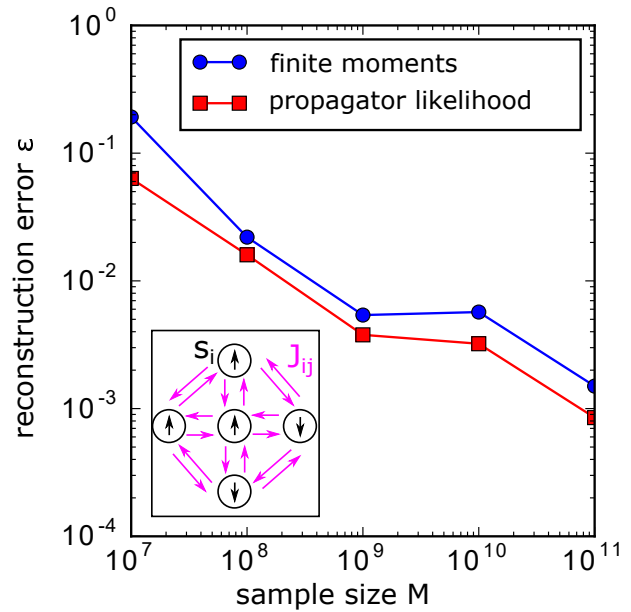}
\caption{The inference of couplings in the kinetic Ising model. The inset schematically shows a system of binary spins interacting via couplings $J_{ij}$ subject to external fields $h_i$ (not shown). In the main figure, we plot the relative error of couplings $\epsilon=\| \mathbf{J}^\text{inf}-\mathbf{J}^{\text{true}}\|_2/\|\mathbf{J}^{\text{true}}\|_2$ versus the number of independent samples used for inference, using (i) finite spin moments up to three-point correlations ($\bullet$) and (ii) the single-step propagator likelihood ($\blacksquare$). Both methods are exact, so the relative error decreases with the sample size as $\epsilon\sim M^{-1/2}$. The propagator likelihood (which uses the full set of configurations sampled) performs only a little better than the fit to the first three moments, showing that most information required for reconstruction is already contained in the first three moments. The underlying off-diagonal couplings were drawn independently from a Gaussian distribution with mean $0$ and standard deviation $1/\sqrt{N}$ (we excluded self-interactions, $J_{ii}= 0$), the external fields were drawn independently from a Gaussian distribution with mean $0$ and standard deviation $1$. The system size was $N=10$ spins. }
\label{fig:ising}
\end{figure}

The kinetic Ising model consists of a set of $N$ binary spins $s_i=\pm1$, which interact with each other via couplings $J_{ij}$ and are subject to external fields $h_i$~(see inset of Fig.~\ref{fig:ising}). Crucially, the couplings are not symmetric ($J_{ij} \neq J_{ji}$ in general). A stochastic dynamics of this model is specified by the so-called Glauber dynamics~\cite{glauber1963a}: In each time step, a spin $i$ is chosen in and its value $s_i(t+1)$ one time step later is updated according to the probability distribution
\begin{equation}
\label{eq:glaubersq}
p(\s_i(t+1)|\sss(t))=\frac{\exp\{\s_i(t+1) \theta_i(t)\}}{2
  \cosh(\theta_i(t))} \ ,
\end{equation}
where the effective local field at time $t$ is 
\begin{equation}
\label{eq:localfield}
\theta_i(t)=h_i+\sum_{j=1}^N J_{ij} \s_j(t)  \ .
\end{equation}
The kinetic Ising model has been used to model gene regulatory and neural networks~\cite{Derrida1987a,hertz1991a,Bailly2010b}.

For a symmetric coupling matrix without self-couplings, the Glauber dynamics~\eqref{eq:glaubersq} converges to the equilibrium state characterised by the Boltzmann distribution ${p_B(\mathbf{s})=e^{-\mathcal{H}(\mathbf{s})}/Z}$ with the well-known Ising Hamiltonian ${\mathcal{H}(\mathbf{s})=-\sum_i s_i (h_i+\sum_{j>i} J_{ij}s_j)}$. For asymmetric couplings, however, Glauber dynamics~\eqref{eq:glaubersq} converges to a non-equilibrium steady state, which lacks detailed balance and is hard to characterise.

In recent work we have shown how the spin couplings $J_{ij}$ and external fields $h_i$ can be inferred from independent samples taken from the steady state by fitting couplings and fields to match the magnetisations, two-, and three-point correlations sampled in the data~\cite{Dettmer2016}. Here we demonstrate that the couplings can be inferred even more accurately with the propagator likelihood~\eqref{eq:propagator-likelihood-def}, which uses information from the full empirical distribution. We insert the single-step propagator~\eqref{eq:glaubersq} into the propagator likelihood~\eqref{eq:propagator-likelihood-def} and maximise the propagator likelihood with respect to the external fields $h_i$ and off-diagonal couplings $J_{ij}$ (we consider a model without self-interactions: $J_{ii}= 0$). For the last step, we use the Broyden-Fletcher-Goldfarb-Shanno algorithm as implemented in the SciPy library~\cite{SciPy}, and initialise the algorithm with the naive mean-field parameter estimate as described in~\cite{Dettmer2016}. Fig.~\ref{fig:ising} compares the relative error of coupling reconstruction $\epsilon=\| \mathbf{J}^\text{inf}-\mathbf{J}^{\text{true}}\|_2/\|\mathbf{J}^{\text{true}}\|_2$ based on the single-step propagator likelihood with the corresponding reconstruction error of fitting finite spin moments up to three-point correlations. 
 
It turns out that parameter inference in the kinetic Ising model requires more samples than in the equilibrium inverse Ising problem. To achieve a relative reconstruction error of $10^{-2}$ for an equilibrium system of $N=10$ spins, the pseudolikelihood method requires of the order of $10^6$ samples~\cite{Aurell2012b}. In the non-equilibrium model considered here, we require at least $10^8$ independent samples for a similar reconstruction accuracy (see Fig.~\ref{fig:ising}). Naturally, inference in the kinetic Ising model becomes significantly easier if time-correlated data is available. For example, the Gaussian mean-field theory~\cite{Mezard2011a} requires only on the order of $10^6$ pairs of samples $\{\sss(t),\sss(t+1)\}$ to achieve a similar reconstruction accuracy for a system as large as 100 spins.
The reason for this is that, in the kinetic Ising model, couplings are not uniquely determined by pairwise correlations. Instead, many different models can reproduce the same pairwise correlations. For this reason, we need information from higher order spin correlations, which require more samples to determine them accurately. 

\begin{figure}[hbtp!]
\includegraphics[width=0.45\textwidth]{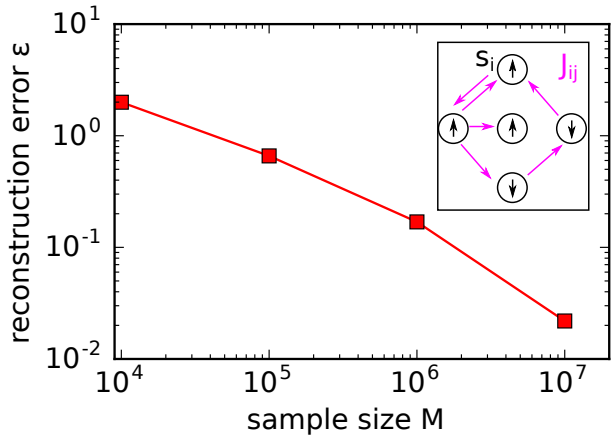}
\caption{Coupling inference in the sparse kinetic Ising model. The inset schematically shows a system of binary spins interacting via sparse couplings $J_{ij}$ subject to external fields $h_i$ (not shown). In the main figure, we plot the relative error of couplings $\epsilon=\| \mathbf{J}^\text{inf}-\mathbf{J}^{\text{true}}\|_2/\|\mathbf{J}^{\text{true}}\|_2$ versus the number of independent samples. The underlying off-diagonal couplings were chosen sparsely: they were set to zero with probability $1-p=0.8$, and with probability $p=0.2$ were drawn independently from a Gaussian distribution with mean 0 and variance $1/N$ (we excluded self-interactions, $J_{ii}=0$ ). The external fields were drawn independently from a Gaussian distribution with mean 0 and variance 1. The system size was $N=10$ spins. The couplings were inferred by maximising the single-step propagator likelihood over the set of couplings between directly interacting spin pairs $(i,j)$, \textit{i.e.} there is at least one true non-zero coupling between the spin pair, $J^\text{true}_{ij}\neq 0$ or $J^\text{true}_{ji}\neq 0$, regardless of the direction.}
\label{fig:ising_sparse}
\end{figure}

\paragraph{Sparse networks.}
We now consider a particular situation, where the parameter inference requires fewer samples: sparse coupling matrices with known topology of the couplings, so only the values of the couplings are to be reconstructed. Specifically, we look at the kinetic Ising model with sparse couplings (so most interactions are zero) and assume as prior knowledge the pairs $(i,j)$ that have a non-zero coupling between them, \textit{i.e.} $J^\text{true}_{ij}\neq 0$ or $J^\text{true}_{ji}\neq 0$, regardless of the direction of the coupling.
This problem has been addressed for undirected equilibrium systems like Ising models with ferromagnetic or binary couplings~\cite{Bento2009,Aurell2012b}. We apply the propagator likelihood to a network of $N=10$ spins, where each possible directed link $J_{ij}$ from spin $i$ to spin $j$ is non-zero with probability $p=0.2$. The non-zero couplings are again drawn independently from a Gaussian distribution with mean $0$ and variance $1/N$. Self-interactions are excluded and the external fields $h_i$ drawn independently from a Gaussian distribution with mean $0$ and variance $1$. Figure~\ref{fig:ising_sparse} shows that the directed couplings can be inferred with slightly fewer samples when the topology of the couplings is known.

%------------------------
\paragraph{Increasing the propagation time interval.}
So far we have restricted ourselves to the single-step propagator ($\tau=1$).  Can the inference be improved by increasing the propagation time interval? Intuitively, we expect that the single-step propagator cannot be improved on when all configurations have been sampled, since this implies that all transitions over longer propagation time intervals consist of single-step transitions that have already been probed by the single-step propagator likelihood: $x^\nu\overset{\tau}{\rightarrow}x^\mu=\sum_{x_1,x_2,...,x_{\tau-1}}x^\nu\overset{\tau=1}{\rightarrow}{x_1}\overset{\tau=1}{\rightarrow}{x_2}\hdots\overset{\tau=1}{\rightarrow}{x_{\tau-1}} \overset{\tau=1}{\rightarrow} x^\mu$. Indeed, the examples with discrete time considered so far in this article fall into this category and our numerical evidence confirms that increasing the propagation time interval does not improve the inference. 
If, however, the configuration space is undersampled, some of the transitions appearing in the longer-time propagator likelihood will involve intermediate configurations that are not present in the sample and therefore do not appear in the single-step propagator likelihood. In this case, we expect to find that increasing the propagation time interval improves the inference for a fixed sample size.
In principle, one could even compute the log-likelihood~\eqref{eq:log-likelihood} numerically by using sufficiently long propagation time intervals $\tau$. However, the computational cost of taking the $2^N$-dimensional transition matrix to a large power $\tau$ is often prohibitive. Furthermore, the matrix products needs to be computed many times in order to evaluate the likelihood and its ($N^2$-dimensional) gradient over many iterations of a maximisation algorithm.

\begin{figure*}
\includegraphics[width=0.9\textwidth]{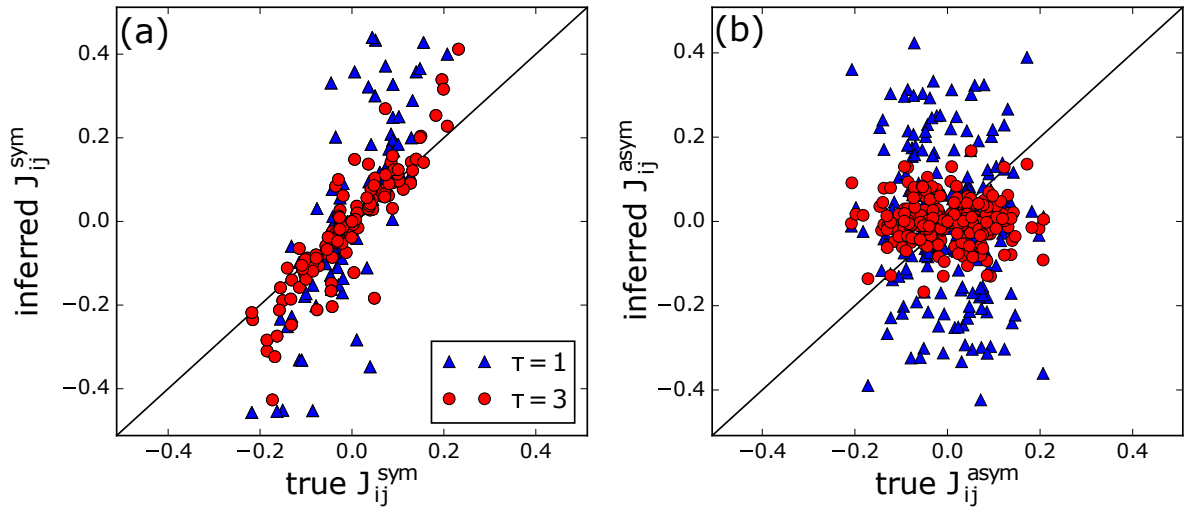}
\caption{Increasing the propagation time interval in the undersampled kinetic Ising model. 
(a) shows the reconstructed symmetric part of the coupling matrix $J^\text{sym}_{ij}=(J_{ij}+J_{ji})/2$ based on the single-step propagator likelihood ($\blacktriangle$) and on the longer propagation time interval $\tau=3$ ($\bullet$). (b) shows the reconstructed antisymmetric part of the coupling matrix $J^\text{asym}_{ij}=(J_{ij}-J_{ji})/2$ also based on the single-step propagator likelihood ($\blacktriangle$) and on the longer propagation time interval $\tau=3$ ($\bullet$). \newline
The underlying off-diagonal couplings were drawn independently from a Gaussian distribution with mean $0$ and standard deviation $0.5/\sqrt{N}$ (we excluded self-interactions, $J_{ii}= 0$), the external fields were drawn independently from a Gaussian distribution with mean $0$ and standard deviation $0.5$. The system size was $N=16$ spins and $M=10^{4}N$ samples were used. As a result, less than a third of the $2^{16}$ system configurations were present in the sample.}
 \label{fig:ising-propagation-time}
\end{figure*}

In Fig.~\ref{fig:ising-propagation-time} we consider a kinetic Ising model where only a small fraction of system configurations appear in the sampled configurations. Increasing the propagation time interval from $\tau=1$ to $\tau=3$ improves the inference markedly. Also, we find that the reconstruction error is much smaller for the symmetric part of the coupling matrix (shown in Fig.~\ref{fig:ising-propagation-time}(a)) than for the antisymmetric part (shown in Fig.~\ref{fig:ising-propagation-time}(b)). This is because the symmetric part of the couplings is governed by the pairwise spin-correlations, while the antisymmetric part is dominated by higher-order spin-correlations, which require more samples for an accurate computation, see~\cite{Dettmer2016}. The benefit of increasing the propagation time interval is also larger for the symmetric part, suggesting that the reconstruction of the antisymmetric part of the couplings is mainly limited by the sample size and that increasing the propagation time interval even further will not lead to a more accurate reconstruction.

%------------------------------------------------------------------------------------------------------------------------------------------------------------
\subsection{The replicator model}

The replicator model describes a dynamics of self-replicating entities, for instance genotypes, different animal species, RNA-molecules, or an abstract strategy in the game-theoretic problem. The replicator model has been used in population genetics, ecology, prebiotic chemistry, and sociobiology~\cite{Schuster1983}.
We consider a population consisting of $N$ different species and denote by $x_i$ the fraction of species $i$ in the total population (scaled for convenience by a factor on $N$ so $\sum_i x_i=N$). The growth rate of species $i$, called its fitness, is denoted by $f_i$. The population fraction change in time depends on the growth rate $f_i$ and the average growth rate of the population $\overline{f}$
\begin{equation}
\frac{\d x_i}{\d t}=x_i(t) (f_i(\mathbf{x},t)-\overline{f}(\mathbf{x},t)) \ ,
\label{eq:replicators-basic}
\end{equation}
with $\overline{f}(\mathbf{x},t)=\frac{1}{N}\sum_{j=1}^N x_j(t)f_j(\mathbf{x},t)$. The set of equations~\eqref{eq:replicators-basic} defines the replicator model.  
The average fitness $\overline{f}$ enters to ensure that the fractions remain normalised such that $\sum_{i} x_i(t)= N$ for all times. 

Here we consider a fitness which for each species $i$ depends linearly on the population fractions of the other species
\begin{equation}
f_i(\mathbf{x}(t))=\sum_{j\neq i}^N J_{ij} x_j(t) \ .
\label{eq:replicators-fitness}
\end{equation}
The inter-species interactions $J_{ij}$ are quenched random variables with mean $u$ (called the cooperation pressure) and standard deviation $1/\sqrt{N}$. There are no self-interactions, $J_{ii}=0$. 

For symmetric interactions, $J_{ij}=J_{ji}$, the fitness vector can be written as the gradient of a Lyapunov function. This implies that the system converges to an equilibrium steady state, which can be characterised by methods from statistical physics~\cite{Diederich1989}. In the socio-biological context, however, there is no reason for the interactions to be symmetric, or in fact to assume deterministic dynamics.
Assuming an asymmetric matrix $J_{ij}$ and allowing random fluctuations $\sigma \xi_i(t)$ in the reproduction of species $i$ leads to a set of Langevin equations 
\begin{equation}
\frac{\d x_i}{\d t}=x_i(t)\left(f_i(\mathbf{x}(t))+\sigma \xi_i(t) -\lambda(\mathbf{x},t)\right) \ ,
\label{eq:replicators-Langevin}
\end{equation}
where the $\xi_i(t)$ are $N$ independent sources of white noise interpreted in the Stratonovich convention, the parameter $\sigma>0$ controls the overall noise strength, and the factor $\lambda(\mathbf{x}(t),t)=\frac{1}{N}\sum_j x_j(t)(f_j(\mathbf{x}(t))+\sigma \xi_j(t))$ ensures normalisation, \textit{i.e.} $\sum_i x_i(t)= N$ for all times.
This dynamics converges to a non-equilibrium steady state. Its characteristics for typical realisations of the matrix of couplings have been studied  in the limit of a large number of species using dynamical mean field theory~\cite{Opper1992}.

\begin{figure*}[hbtp!]
\includegraphics[width=0.9\textwidth]{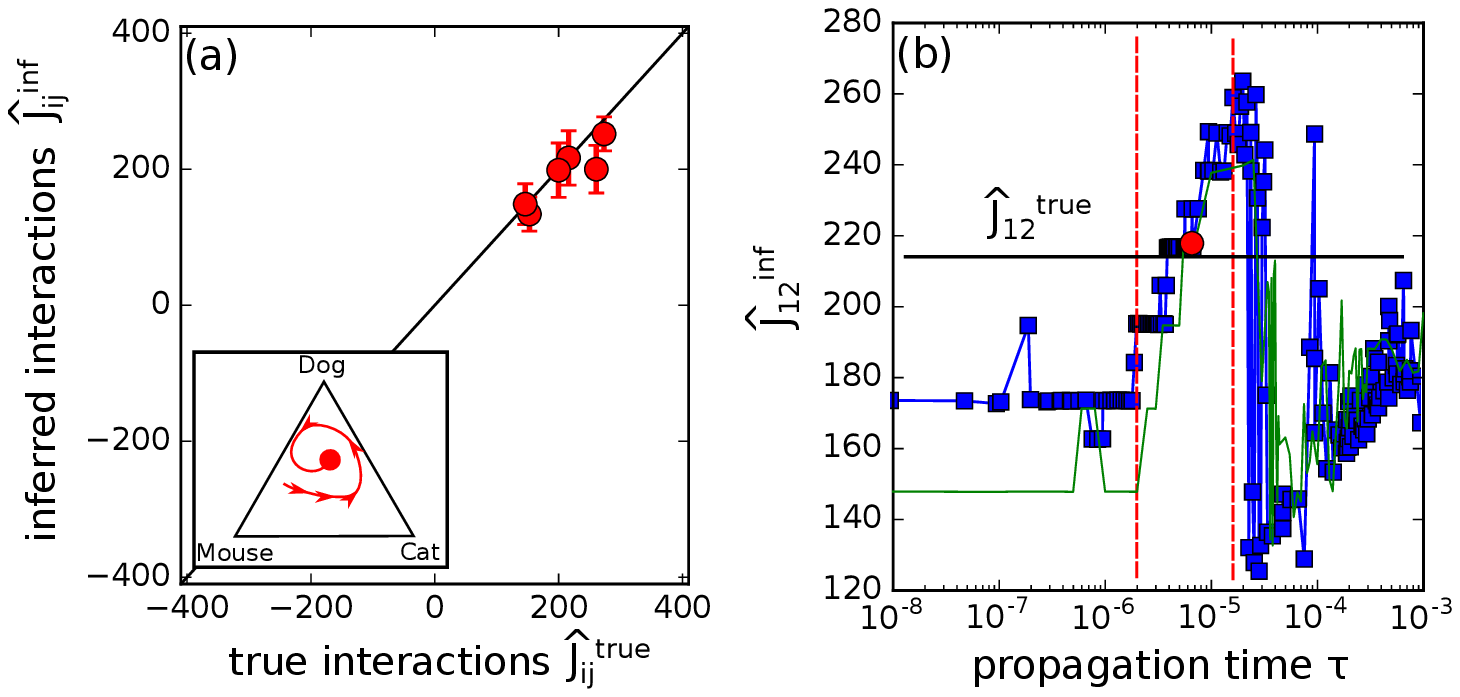}
\caption{Reconstruction of the inter-species interactions in replicator dynamics. \newline (a) The inset schematically shows the replicator model describing the population dynamics of different species competing for fractions of the total population size. The population moves on a $N-1$-dimensional simplex defined by the normalisation $\sum_i x_i=N,\  x_i \geq 0$. In the main figure, we plot the inferred rescaled inter-species interactions $\hat{J}^\text{inf}_{ij}\equiv J^\text{inf}_{ij}/\sigma^2$ versus the rescaled underlying interactions $\hat{J}^\text{true}_{ij}= {J}^\text{true}_{ij}/\sigma^2$ for the propagation time interval $\tau=5.0\times 10^{-6}$. The error bars indicate the error due to the ambiguity associated with the choice of the propagation time interval $\tau$ as described next. (b) shows how the propagation time interval was chosen and how the reconstruction error can be estimated without recourse to the underlying couplings. For this plot, an arbitrary parameter (here $\hat{J}_{12}$) was chosen and its inferred value plotted for different propagation time intervals $\tau_i$ ($\blacksquare$). The horizontal line shows the value of the true underlying parameter $\hat{J}_{12}^\text{true}$.
For small values of $\tau$ the effects of sampling fluctuations dominate and the inferred parameter saturates as discussed in section~\ref{sec:OUP}.
For large $\tau$, the error due to the linearisation of the Langevin equations is large and the inference becomes unstable, as signalled by the erratic changes in the value of the inferred parameter. 
The interval of reasonable propagation time intervals must lie between those two regimes and we choose a propagation time (marked by the circle) that lies in the (logarithmic) centre of this transition region (marked by the two vertical dashed lines). The other parameters show a similar behaviour and the same transition time interval, so the choice of the propagation time interval does not depend on the parameter considered. 
For each parameter, we take the vertical extent of the transition region as the estimation error. To illustrate the effects of sampling fluctuations, we repeated the procedure above a second time with the same model parameters but different samples (continuous line without markers). As expected, the sampling fluctuations influence mainly the inferred parameters for $\tau \rightarrow 0$, while the inference for larger values of $\tau$ is far less sensitive to the fluctuations.\newline
The system consisted of $N=3$ species, the noise strength was set to $\sigma=0.1$, and the underlying interactions $J^\text{true}_{ij}$ were quenched random variables chosen independently from a Gaussian with mean $u=2.0$ and standard deviation $1/\sqrt{N}$ (no self-interactions: $J_{ii}= 0$).
We used an Euler discretisation of the Langevin equation~\eqref{eq:replicators-Langevin} with time steps of length $\Delta t=10^{-6}/\sigma^2$ and a total of $M=10^4$ samples were taken every $10^4$ steps after an initial settling time of $10^9$ steps.}
\label{fig:replicators}
\end{figure*}

We now turn to the problem of inferring the couplings $J_{ij}$ of the replicator model from a set of configurations $\{\mathbf{x}^\mu\}_{\mu=1}^M$ taken independently from the non-equilibrium steady state. For simplicity, we focus on the so-called cooperative regime, in which all species survive in the long-time limit, \textit{i.e.} $\lim_{t\rightarrow \infty}x_i(t)>0 \ \forall i$. 
This regime is characterised by a sufficiently large value of the cooperation pressure $u$~\cite{Opper1992}.
Our results can be generalised to the case where species go extinct by restricting the transitions $\mathbf{x}^\nu \rightarrow \mathbf{x}^\mu$ considered in the propagator likelihood to those between configurations with the same set of surviving species. 

Again, to make time dimensionless, we rescale time $t'=t\sigma^2$, resulting in a noise-term with unit magnitude. The steady state and the propagator depend only on the rescaled couplings $\hat{J}_{ij} \equiv J_{ij}/\sigma^2$.
By linearising the LE~\eqref{eq:replicators-Langevin} for short times and eliminating $x_N$  via the normalisation constraint,  $x_N=N-\sum_{i=1}^{N-1}x_i$, we arrive at the Gaussian short-term propagator
\begin{align}
&p(\mathbf{x},\tau|\mathbf{y},0) \approx\frac{1}{\sqrt{2\pi \tau}^{N-1}\sqrt{\text{Det} \Sigma}} \times  \notag \\
&\exp \left\{ -\frac{1}{2 \tau}\sum_{i,j=1}^{N-1}\left(x_i-y_i-\mu_i\tau\right) \Sigma^{-1}_{ij}  \left(x_j-y_j-\mu_j\tau\right) \right\}
\label{eq:replicator-propagator}
\end{align}
with drift~\footnote{The second term in the drift arises from the difference between the It{\^o} and Stratonovich convention in the Langevin equation.}
\begin{align}
\mu_i &= y_i(\hat{f}_i(\mathbf{y})-\overline{\hat{f}}(\mathbf{y}))-\frac{y_i}{N}\left(y_i-\frac{1}{N}\sum_{j=1}^N y_j^2\right)  \notag \\
\label{eq:propagator-drift}
\end{align}
and covariance matrix $\Sigma=A A^T\in \mathbb{R}^{N-1\times N-1}$ with
\begin{equation}
A_{ij}=y_i(y_j/N-\delta_{i,j}) \ .
\label{eq:propagator-covariance}
\end{equation}
We denote by $\hat{f}_i(\mathbf{y})$ the fitness~\eqref{eq:replicators-fitness} calculated with the rescaled variables $\hat{J}_{ij}=J_{ij}/\sigma^2$, instead of the original interactions $J_{ij}$, and by $\overline{\hat{f}}(\mathbf{y})=\frac{1}{N}\sum_j y_j\hat{f}_j(\mathbf{y})$ its species-weighted average. 

To reconstruct the rescaled interactions $\hat{J}_{ij}$, we insert the Gaussian short-term propagator~\eqref{eq:replicator-propagator} into the propagator likelihood~\eqref{eq:propagator-likelihood-def} and maximise it using the Broyden-Fletcher-Goldfarb-Shanno algorithm (see Fig.~\ref{fig:replicators}). As for the OUP, the reconstruction error depends non-monotonically on the choice of the dimensionless propagation time interval $\tau$, due to the trade-off between the error from linearising the LE and the error from effectively reducing the sample size by exponentially damping the propagators of most transitions. Unfortunately, the simple procedure we used for the OUP, minimising the parameter derivative $|\partial \Theta^\text{inf}/\partial \tau|$, cannot easily be generalised to higher dimensions. The reason is that the derivative of the reconstruction error $\partial \epsilon /\partial \tau$ is a linear combination of the individual parameter entries $(\partial \Theta^\text
 {inf}_i/\partial \tau)_{i=1}^K$, which can cancel each other without vanishing individually (here $K=N(N-1)$ denotes the number of model parameters). To see that not all individual derivatives can vanish simultaneously, we remind ourselves that the inferred parameters must satisfy $0\equiv \frac{\partial \mathcal{PL}}{\partial \Theta_i} (\Theta^\text{inf}(\tau),\tau) \ ,i=1,\hdots,K$. Additionally demanding $ \partial \Theta^\text{inf}_i/\partial \tau=0 ,i=1,\hdots,K$, corresponds to solving the system of equations $\{\frac{\partial \mathcal{PL}}{\partial \Theta_i}=0,\frac{\partial^2 \mathcal{PL}}{\partial \Theta_i \partial \tau}=0 \}_{i=1}^{K}$ for the $K+1$ variables $(\Theta_i,\tau)$. This system of $2K$ nonlinear equations for $K+1$ variables will in general have no solution for $K>1$. Instead, we can find a good propagation time interval by plotting a single inferred parameter versus the propagation time interval $\tau$ used for inference [see Fig.~\ref{fig:replicators}(b)]. The regime where the inference is dominated by the error from the linearisation for large values of $\tau$ is characterised by an erratic change of the value of the inferred parameter. At small values of $\tau$, the reconstruction is dominated by sampling fluctuations (see section~\ref{sec:OUP}). These regimes are connected by a transition region, from which the propagation time interval should be chosen. We checked that this transition region stretched across the same time interval (approximately $[2\times 10^5,2\times 10^6]$) for all parameters and chose the logarithmic center of this transition interval as the propagation time interval $\tau$. We found this produced a good reconstruction quality, however, a method to pinpoint the optimal value of $\tau$ is currently lacking.

%===========================================================
\FloatBarrier
\section{Conclusions}
We study parameter inference for a non-equilibrium model from independent samples taken from the steady state. Our approach is based on a variant of the likelihood we call the propagator likelihood. In the limit of a large propagation time interval, the propagator likelihood converges to the likelihood of the model. However, for non-equilibrium system, the likelihood and the limit of large propagation time intervals is generally intractable. The propagator likelihood can be derived from a variational principle aiming to find model parameters for which the distribution of configurations sampled from the steady state is invariant under propagation in time.

For systems with discrete configurations, we base our reconstruction on the single-step propagator, although increasing the propagation time interval can improve the inference when not all configurations have been sampled. This can be understood as follows: at short times, most pairs of sampled configurations have a small or even vanishing propagator, and the propagator likelihood~\eqref{eq:propagator-likelihood-def} is dominated by a few pairs of close configurations. At higher values of the propagation time interval $\tau$, more configuration pairs contribute to the propagator likelihood, which reduces sampling fluctuations. However, as the computational complexity of evaluating the propagator grows exponentially with the number of time steps, there is a competition between inference quality and computational complexity. For systems with continuous configurations, we use a short-time approximation to the propagator. Also in this case, inference improves with the propagation time interval $\tau$ until the short-time approximation becomes invalid. \newline

Inferring model parameters from the steady state requires a large number of samples: Inferring couplings of the kinetic Ising model with $N=10$ spins to within a reconstruction error $\epsilon \approx 0.01$ requires $M \approx 10^8$ samples, compared to the equilibrium case requiring approximately $10^6$ samples (for couplings drawn independently from a Gaussian with mean 0 and variance $1/N$). The bottleneck in practical applications may thus well be the number of available samples. 
Non-equilibrium inference is also computationally expensive: evaluating the propagator likelihood takes $\mathcal{O}(M^2)$ operations for systems with continuous configurations and $\mathcal{O}(M)$ operations for systems with discrete configurations (provided that only a small number of neighbouring configurations can be reached in a single step with non-zero transition probability). A challenge for the future is to find more efficient inference methods, both in terms of the number of samples required and in terms of the computational complexity.

\acknowledgments{This work was supported by the BMBF [grant number emed:SMOOSE].}


\begin{thebibliography}{20}%
\makeatletter
\providecommand \@ifxundefined [1]{%
 \@ifx{#1\undefined}
}%
\providecommand \@ifnum [1]{%
 \ifnum #1\expandafter \@firstoftwo
 \else \expandafter \@secondoftwo
 \fi
}%
\providecommand \@ifx [1]{%
 \ifx #1\expandafter \@firstoftwo
 \else \expandafter \@secondoftwo
 \fi
}%
\providecommand \natexlab [1]{#1}%
\providecommand \enquote  [1]{``#1''}%
\providecommand \bibnamefont  [1]{#1}%
\providecommand \bibfnamefont [1]{#1}%
\providecommand \citenamefont [1]{#1}%
\providecommand \href@noop [0]{\@secondoftwo}%
\providecommand \href [0]{\begingroup \@sanitize@url \@href}%
\providecommand \@href[1]{\@@startlink{#1}\@@href}%
\providecommand \@@href[1]{\endgroup#1\@@endlink}%
\providecommand \@sanitize@url [0]{\catcode `\\12\catcode `\$12\catcode
  `\&12\catcode `\#12\catcode `\^12\catcode `\_12\catcode `\%12\relax}%
\providecommand \@@startlink[1]{}%
\providecommand \@@endlink[0]{}%
\providecommand \url  [0]{\begingroup\@sanitize@url \@url }%
\providecommand \@url [1]{\endgroup\@href {#1}{\urlprefix }}%
\providecommand \urlprefix  [0]{URL }%
\providecommand \Eprint [0]{\href }%
\providecommand \doibase [0]{http://dx.doi.org/}%
\providecommand \selectlanguage [0]{\@gobble}%
\providecommand \bibinfo  [0]{\@secondoftwo}%
\providecommand \bibfield  [0]{\@secondoftwo}%
\providecommand \translation [1]{[#1]}%
\providecommand \BibitemOpen [0]{}%
\providecommand \bibitemStop [0]{}%
\providecommand \bibitemNoStop [0]{.\EOS\space}%
\providecommand \EOS [0]{\spacefactor3000\relax}%
\providecommand \BibitemShut  [1]{\csname bibitem#1\endcsname}%
\let\auto@bib@innerbib\@empty
%</preamble>
\bibitem [{\citenamefont {Nguyen}\ \emph {et~al.}(2017)\citenamefont {Nguyen},
  \citenamefont {Zecchina},\ and\ \citenamefont {Berg}}]{Nguyen2017}%
  \BibitemOpen
  \bibfield  {author} {\bibinfo {author} {\bibfnamefont {H.~C.}\ \bibnamefont
  {Nguyen}}, \bibinfo {author} {\bibfnamefont {R.}~\bibnamefont {Zecchina}}, \
  and\ \bibinfo {author} {\bibfnamefont {J.}~\bibnamefont {Berg}},\ }\href {\doibase 10.1080/00018732.2017.1341604} {\bibfield  {journal} {\bibinfo  {journal} {Adv. Phys.}\ }\textbf {\bibinfo {volume} {66}} (\bibinfo {number} {3}),  \bibinfo {pages} {197-261}  ,\ (\bibinfo {year}
  {2017})}\BibitemShut {NoStop}%
\bibitem [{\citenamefont {Roudi}\ and\ \citenamefont
  {Hertz}(2011)}]{Roudi2011a}%
  \BibitemOpen
  \bibfield  {author} {\bibinfo {author} {\bibfnamefont {Y.}~\bibnamefont
  {Roudi}}\ and\ \bibinfo {author} {\bibfnamefont {J.}~\bibnamefont {Hertz}},\
  }\href {\doibase 10.1103/PhysRevLett.106.048702} {\bibfield  {journal}
  {\bibinfo  {journal} {Phys. Rev. Lett.}\ }\textbf {\bibinfo {volume} {106}},\
  \bibinfo {pages} {048702} (\bibinfo {year} {2011})}\BibitemShut {NoStop}%
\bibitem [{\citenamefont {Zeng}\ \emph {et~al.}(2013)\citenamefont {Zeng},
  \citenamefont {Alava}, \citenamefont {Aurell}, \citenamefont {Hertz},\ and\
  \citenamefont {Roudi}}]{zeng2013}%
  \BibitemOpen
  \bibfield  {author} {\bibinfo {author} {\bibfnamefont {H.-L.}\ \bibnamefont
  {Zeng}}, \bibinfo {author} {\bibfnamefont {M.}~\bibnamefont {Alava}},
  \bibinfo {author} {\bibfnamefont {E.}~\bibnamefont {Aurell}}, \bibinfo
  {author} {\bibfnamefont {J.}~\bibnamefont {Hertz}}, \ and\ \bibinfo {author}
  {\bibfnamefont {Y.}~\bibnamefont {Roudi}},\ }\href {\doibase
  10.1103/PhysRevLett.110.210601} {\bibfield  {journal} {\bibinfo  {journal}
  {Phys. Rev. Lett.}\ }\textbf {\bibinfo {volume} {110}},\ \bibinfo {pages}
  {210601} (\bibinfo {year} {2013})}\BibitemShut {NoStop}%
\bibitem [{\citenamefont {M\'{e}zard}\ and\ \citenamefont
  {Sakellariou}(2011)}]{Mezard2011a}%
  \BibitemOpen
  \bibfield  {author} {\bibinfo {author} {\bibfnamefont {M.}~\bibnamefont
  {M\'{e}zard}}\ and\ \bibinfo {author} {\bibfnamefont {J.}~\bibnamefont
  {Sakellariou}},\ }\href {\doibase 10.1088/1742-5468/2011/07/L07001}
  {\bibfield  {journal} {\bibinfo  {journal} {J. Stat. Mech.}\ ,\ \bibinfo
  {pages} {L07001}} (\bibinfo {year} {2011})}\BibitemShut {NoStop}%
\bibitem [{\citenamefont {Mahmoudi}\ and\ \citenamefont
  {Saad}(2014)}]{Mahmoudi2014}%
  \BibitemOpen
  \bibfield  {author} {\bibinfo {author} {\bibfnamefont {H.}~\bibnamefont
  {Mahmoudi}}\ and\ \bibinfo {author} {\bibfnamefont {D.}~\bibnamefont
  {Saad}},\ }\href {\doibase 10.1088/1742-5468/2014/07/P07001} {\bibfield
  {journal} {\bibinfo  {journal} {{J. Stat. Mech.}}\ ,\ \bibinfo {pages}
  {P07001}} (\bibinfo {year} {2014})}\BibitemShut {NoStop}%
\bibitem [{\citenamefont {Kullback}\ and\ \citenamefont
  {Leibler}(1951)}]{Kullback1951}%
  \BibitemOpen
  \bibfield  {author} {\bibinfo {author} {\bibfnamefont {S.}~\bibnamefont
  {Kullback}}\ and\ \bibinfo {author} {\bibfnamefont {R.~A.}\ \bibnamefont
  {Leibler}},\ }\href {\doibase 10.1214/aoms/1177729694} {\bibfield  {journal}
  {\bibinfo  {journal} {Ann. Math. Statist.}\ }\textbf {\bibinfo {volume}
  {22}},\ \bibinfo {pages} {79} (\bibinfo {year} {1951})}\BibitemShut {NoStop}%
\bibitem [{\citenamefont {Krug}\ and\ \citenamefont
  {Ferrari}(1996)}]{Krug1996}%
  \BibitemOpen
  \bibfield  {author} {\bibinfo {author} {\bibfnamefont {J.}~\bibnamefont
  {Krug}}\ and\ \bibinfo {author} {\bibfnamefont {P.}~\bibnamefont {Ferrari}},\
  }\href {\doibase 10.1088/0305-4470/29/18/004} {\bibfield  {journal} {\bibinfo
   {journal} {J. Phys. A-Math. Gen.}\ }\textbf {\bibinfo {volume} {29}},\
  \bibinfo {pages} {L465} (\bibinfo {year} {1996})}\BibitemShut {NoStop}%
\bibitem [{\citenamefont {Evans}(1997)}]{Evans1997}%
  \BibitemOpen
  \bibfield  {author} {\bibinfo {author} {\bibfnamefont {M.}~\bibnamefont
  {Evans}},\ }\href {\doibase 10.1088/0305-4470/30/16/011} {\bibfield
  {journal} {\bibinfo  {journal} {J. Phys. A-Math. Gen.}\ }\textbf {\bibinfo
  {volume} {30}},\ \bibinfo {pages} {5669} (\bibinfo {year}
  {1997})}\BibitemShut {NoStop}%
\bibitem [{\citenamefont {Derrida}(1998)}]{Derrida1998}%
  \BibitemOpen
  \bibfield  {author} {\bibinfo {author} {\bibfnamefont {B.}~\bibnamefont
  {Derrida}},\ }\href {\doibase 10.1016/S0370-1573(98)00006-4} {\bibfield
  {journal} {\bibinfo  {journal} {Phys. Rep.}\ }\textbf {\bibinfo {volume}
  {301}},\ \bibinfo {pages} {65} (\bibinfo {year} {1998})}\BibitemShut
  {NoStop}%
\bibitem [{\citenamefont {Jones}\ \emph {et~al.}(01  )\citenamefont {Jones},
  \citenamefont {Oliphant}, \citenamefont {Peterson} \emph {et~al.}}]{SciPy}%
  \BibitemOpen
  \bibfield  {author} {\bibinfo {author} {\bibfnamefont {E.}~\bibnamefont
  {Jones}}, \bibinfo {author} {\bibfnamefont {T.}~\bibnamefont {Oliphant}},
  \bibinfo {author} {\bibfnamefont {P.}~\bibnamefont {Peterson}},  \emph
  {et~al.},\ }\href {http://www.scipy.org/} { {\bibinfo {title}
  {{SciPy}: Open source scientific tools for {Python}},}\ } (\bibinfo {year}
  {2001--}),\ \bibinfo {note} {[Online; accessed 2016-07-08]}\BibitemShut
  {NoStop}%
\bibitem [{\citenamefont {Glauber}(1963)}]{glauber1963a}%
  \BibitemOpen
  \bibfield  {author} {\bibinfo {author} {\bibfnamefont {R.~J.}\ \bibnamefont
  {Glauber}},\ }\href {\doibase 10.1063/1.1703954} {\bibfield  {journal}
  {\bibinfo  {journal} {J. Math. Phys.}\ }\textbf {\bibinfo {volume} {4}},\
  \bibinfo {pages} {294} (\bibinfo {year} {1963})}\BibitemShut {NoStop}%
\bibitem [{\citenamefont {Derrida}\ \emph {et~al.}(1987)\citenamefont
  {Derrida}, \citenamefont {Gardner},\ and\ \citenamefont
  {Zippelius}}]{Derrida1987a}%
  \BibitemOpen
  \bibfield  {author} {\bibinfo {author} {\bibfnamefont {B.}~\bibnamefont
  {Derrida}}, \bibinfo {author} {\bibfnamefont {E.}~\bibnamefont {Gardner}}, \
  and\ \bibinfo {author} {\bibfnamefont {A.}~\bibnamefont {Zippelius}},\ }\href
  {\doibase 10.1209/0295-5075/4/2/007} {\bibfield  {journal} {\bibinfo
  {journal} {Europhys. Lett.}\ }\textbf {\bibinfo {volume} {4}},\ \bibinfo
  {pages} {167} (\bibinfo {year} {1987})}\BibitemShut {NoStop}%
\bibitem [{\citenamefont {Hertz}\ \emph {et~al.}(1991)\citenamefont {Hertz},
  \citenamefont {Krogh},\ and\ \citenamefont {Palmer}}]{hertz1991a}%
  \BibitemOpen
  \bibfield  {author} {\bibinfo {author} {\bibfnamefont {J.}~\bibnamefont
  {Hertz}}, \bibinfo {author} {\bibfnamefont {A.}~\bibnamefont {Krogh}}, \ and\
  \bibinfo {author} {\bibfnamefont {R.~G.}\ \bibnamefont {Palmer}},\
  }\href@noop {} {\emph {\bibinfo {title} {Introduction to the theory of neural
  computation}}},\ Vol.~\bibinfo {volume} {1}\ (\bibinfo  {publisher}
  {Addison-Wesley Publishing Company},\ \bibinfo {year} {1991})\BibitemShut
  {NoStop}%
\bibitem [{\citenamefont {Bailly-Bechet}\ \emph {et~al.}(2010)\citenamefont
  {Bailly-Bechet}, \citenamefont {Braunstein}, \citenamefont {Pagnani},
  \citenamefont {Weigt},\ and\ \citenamefont {Zecchina}}]{Bailly2010b}%
  \BibitemOpen
  \bibfield  {author} {\bibinfo {author} {\bibfnamefont {M.}~\bibnamefont
  {Bailly-Bechet}}, \bibinfo {author} {\bibfnamefont {A.}~\bibnamefont
  {Braunstein}}, \bibinfo {author} {\bibfnamefont {A.}~\bibnamefont {Pagnani}},
  \bibinfo {author} {\bibfnamefont {M.}~\bibnamefont {Weigt}}, \ and\ \bibinfo
  {author} {\bibfnamefont {R.}~\bibnamefont {Zecchina}},\ }\href {\doibase
  10.1186/1471-2105-11-355} {\bibfield  {journal} {\bibinfo  {journal} {BMC
  Bioinformatics}\ }\textbf {\bibinfo {volume} {11}},\ \bibinfo {pages} {355}
  (\bibinfo {year} {2010})}\BibitemShut {NoStop}%
\bibitem [{\citenamefont {Dettmer}\ \emph {et~al.}(2016)\citenamefont
  {Dettmer}, \citenamefont {Nguyen},\ and\ \citenamefont {Berg}}]{Dettmer2016}%
  \BibitemOpen
  \bibfield  {author} {\bibinfo {author} {\bibfnamefont {S.~L.}\ \bibnamefont
  {Dettmer}}, \bibinfo {author} {\bibfnamefont {H.~C.}\ \bibnamefont {Nguyen}},
  \ and\ \bibinfo {author} {\bibfnamefont {J.}~\bibnamefont {Berg}},\ }\href
  {\doibase {10.1103/PhysRevE.94.052116}} {\bibfield  {journal} {\bibinfo
  {journal} {{Phys. Rev. E}}\ }\textbf {\bibinfo {volume} {{94}}},\ \bibinfo
  {pages} {052116} (\bibinfo {year} {{2016}})}\BibitemShut {NoStop}%
\bibitem [{\citenamefont {Aurell}\ and\ \citenamefont
  {Ekeberg}(2012)}]{Aurell2012b}%
  \BibitemOpen
  \bibfield  {author} {\bibinfo {author} {\bibfnamefont {E.}~\bibnamefont
  {Aurell}}\ and\ \bibinfo {author} {\bibfnamefont {M.}~\bibnamefont
  {Ekeberg}},\ }\href {\doibase 10.1103/PhysRevLett.108.090201} {\bibfield
  {journal} {\bibinfo  {journal} {Phys. Rev. Lett.}\ }\textbf {\bibinfo
  {volume} {108}},\ \bibinfo {pages} {090201} (\bibinfo {year}
  {2012})}\BibitemShut {NoStop}%
\bibitem [{\citenamefont {Bento}\ and\ \citenamefont
  {Montanari}(2009)}]{Bento2009}%
  \BibitemOpen
  \bibfield  {author} {\bibinfo {author} {\bibfnamefont {J.}~\bibnamefont
  {Bento}}\ and\ \bibinfo {author} {\bibfnamefont {A.}~\bibnamefont
  {Montanari}},\ }\href@noop {} {\bibfield  {journal} {\bibinfo  {journal}
  {Adv. Neural Inf. Process. Syst.}\ }\textbf {\bibinfo {volume} {22}}
  (\bibinfo {year} {2009})}\BibitemShut {NoStop}%
\bibitem [{\citenamefont {Schuster}\ and\ \citenamefont
  {Sigmund}(1983)}]{Schuster1983}%
  \BibitemOpen
  \bibfield  {author} {\bibinfo {author} {\bibfnamefont {P.}~\bibnamefont
  {Schuster}}\ and\ \bibinfo {author} {\bibfnamefont {K.}~\bibnamefont
  {Sigmund}},\ }\href {\doibase 10.1016/0022-5193(83)90445-9} {\bibfield
  {journal} {\bibinfo  {journal} {J. Theor. Biol.}\ }\textbf {\bibinfo {volume}
  {100}},\ \bibinfo {pages} {533} (\bibinfo {year} {1983})}\BibitemShut
  {NoStop}%
\bibitem [{\citenamefont {Diederich}\ and\ \citenamefont
  {Opper}(1989)}]{Diederich1989}%
  \BibitemOpen
  \bibfield  {author} {\bibinfo {author} {\bibfnamefont {S.}~\bibnamefont
  {Diederich}}\ and\ \bibinfo {author} {\bibfnamefont {M.}~\bibnamefont
  {Opper}},\ }\href {\doibase 10.1103/PhysRevA.39.4333} {\bibfield  {journal}
  {\bibinfo  {journal} {Phys. Rev. A}\ }\textbf {\bibinfo {volume} {39}},\
  \bibinfo {pages} {4333} (\bibinfo {year} {1989})}\BibitemShut {NoStop}%
\bibitem [{\citenamefont {Opper}\ and\ \citenamefont
  {Diederich}(1992)}]{Opper1992}%
  \BibitemOpen
  \bibfield  {author} {\bibinfo {author} {\bibfnamefont {M.}~\bibnamefont
  {Opper}}\ and\ \bibinfo {author} {\bibfnamefont {S.}~\bibnamefont
  {Diederich}},\ }\href {\doibase 10.1103/PhysRevLett.69.1616} {\bibfield
  {journal} {\bibinfo  {journal} {Phys. Rev. Lett.}\ }\textbf {\bibinfo
  {volume} {69}},\ \bibinfo {pages} {1616} (\bibinfo {year}
  {1992})}\BibitemShut {NoStop}%
\end{thebibliography}
\end{document}